\documentclass[10pt, twocolumn, nofootinbib,superscriptaddress]{revtex4-2}
\usepackage{amsmath,amssymb,amsfonts}
\usepackage{algorithmic}
\usepackage{graphicx}
\usepackage{textcomp}
\usepackage{xcolor}
\usepackage{ragged2e}
\usepackage{booktabs, makecell, tabularx}
\usepackage{lipsum}

\usepackage{gensymb}
\usepackage{etoolbox}
\usepackage[utf8]{inputenc}

\makeatletter
    \renewcommand\@make@capt@title[2]{%
     \@ifx@empty\float@link{\@firstofone}{\expandafter\href\expandafter{\float@link}}%
      {\textbf{#1}}\@caption@fignum@sep#2\quad}%
\makeatother
\makeatletter 
\renewcommand{\fnum@figure}{\textbf{Fig.~\thefigure}} 
\makeatother

\usepackage{xcolor}

\def\BibTeX{{\rm B\kern-.05em{\sc i\kern-.025em b}\kern-.08em
    T\kern-.1667em\lower.7ex\hbox{E}\kern-.125emX}}

\makeatletter
\renewcommand\normalsize{%
\@setfontsize\normalsize\@xpt\@xiipt
\abovedisplayskip 1\p@ \@plus2\p@ \@minus5\p@
\abovedisplayshortskip \z@ \@plus3\p@
\belowdisplayshortskip 6\p@ \@plus3\p@ \@minus3\p@
\belowdisplayskip \abovedisplayskip
\let\@listi\@listI}
\makeatother

\begin{document}

\author{Shangqing Shi}
\thanks{These authors contributed equally to this work}
\affiliation{Nonlinear Nanophotonics Group, MESA+ Institute of Nanotechnology,\\
University of Twente, Enschede, Netherlands}
\affiliation{Advanced Photonics Center, School of Electronic Science and Engineering,\\
Southeast University, Nanjing, China}

\author{Kaixuan Ye}
\thanks{These authors contributed equally to this work}
\affiliation{Nonlinear Nanophotonics Group, MESA+ Institute of Nanotechnology,\\
University of Twente, Enschede, Netherlands}

\author{Martijn~van~den~Berg}
\affiliation{Nonlinear Nanophotonics Group, MESA+ Institute of Nanotechnology,\\
University of Twente, Enschede, Netherlands}

\author{Okky~Daulay}
\affiliation{Nonlinear Nanophotonics Group, MESA+ Institute of Nanotechnology,\\
University of Twente, Enschede, Netherlands}

\author{Chuangchuang Wei}
\affiliation{Nonlinear Nanophotonics Group, MESA+ Institute of Nanotechnology,\\
University of Twente, Enschede, Netherlands}

\author{Binfeng Yun}
\affiliation{Advanced Photonics Center, School of Electronic Science and Engineering,\\
Southeast University, Nanjing, China}

\author{David Marpaung}
\email{david.marpaung@utwente.nl}
\affiliation{Nonlinear Nanophotonics Group, MESA+ Institute of Nanotechnology,\\
University of Twente, Enschede, Netherlands}

\date{\today}
\title{Integrated RF Photonic Front-End Capable of Simultaneous Cascaded Functions}

\begin{abstract}

Integrated microwave photonic (MWP) front-ends are capable of ultra-broadband signal reception and processing. However, state-of-the-art demonstrations are limited to performing only one specific functionality at any given time, which fails to meet the demands of advanced radio frequency applications in real-world electromagnetic environments. In this paper, we present a major departure from the current trend, which is a novel integrated MWP front-end capable of simultaneous cascaded functions with enhanced performances. Our integrated MWP front-end can delay or phase-shift signals within the selected frequency band while simultaneously suppressing noise signals in other frequency bands, resembling the function of a conventional RF front-end chain. Moreover, we implement an on-chip linearization technique to improve the spurious-free dynamic range of the system. Our work represents a paradigm shift in designing RF photonic front-ends and advancing their practical applications.

\end{abstract}
\maketitle

\section*{Introduction}

The rapid advancement of integrated sensing and communication technologies, along with cognitive, intelligent, and other advanced radio frequency (RF)/microwave applications \cite{r1-10368012,liu2022integrated}, is driving an ever-increasing demand for RF front-ends with enhanced functionality, performance, and integration. To address this evolving requirements, integrated optics and microwave photonic (MWP) technologies have been combined to develop advanced integrated MWP front-ends. These front-ends offer significant advantages, including high-speed transmission, wide bandwidth signal reception and processing, reconfigurability, and immunity to electromagnetic interference \cite{r1-marpaung2019integrated,yao2022microwave,r3-feng2024integrated}. 

\begin{figure}[t!]
\centering
\includegraphics[width=\linewidth]{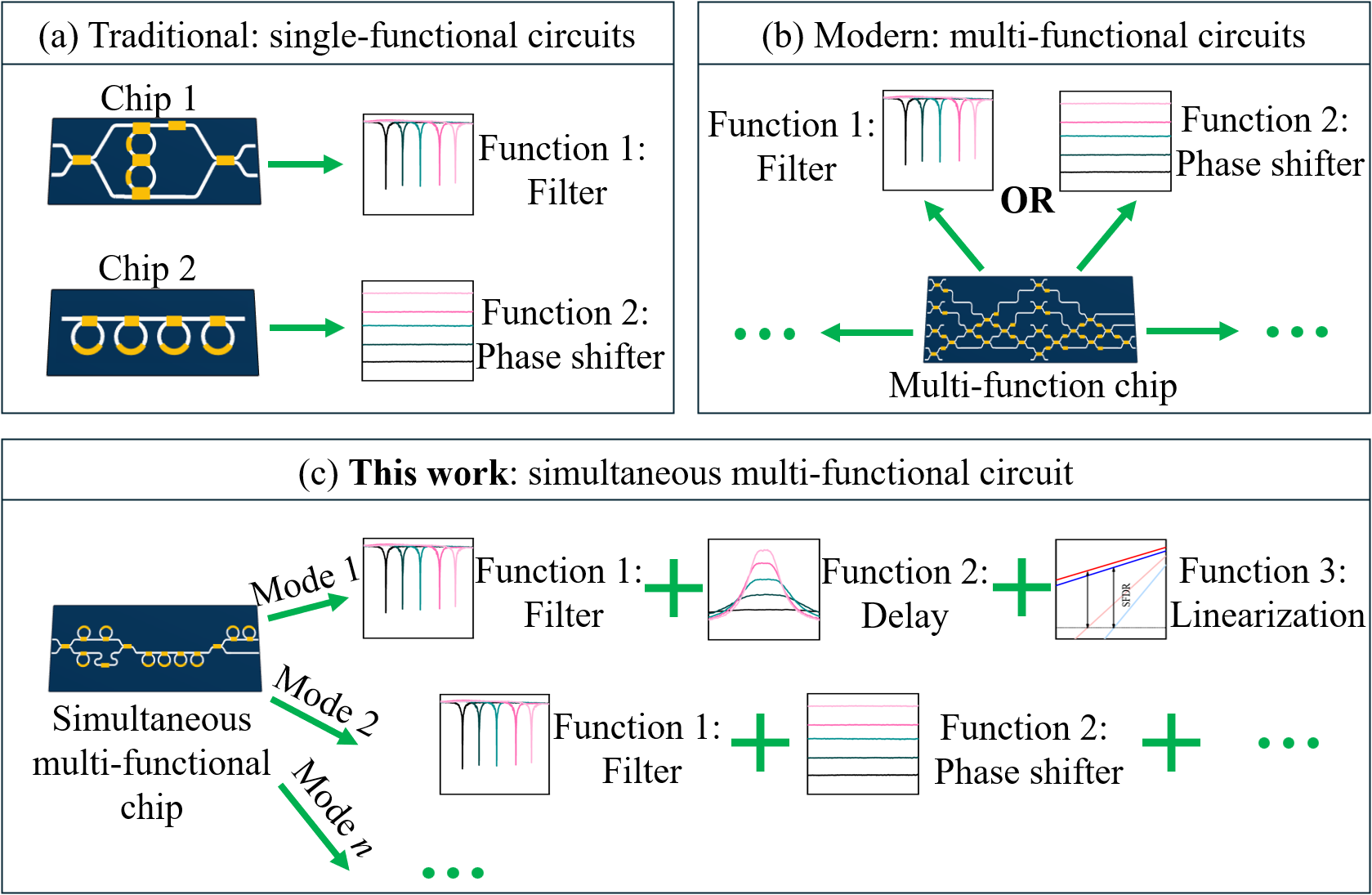}
\caption{Development of integrated RF photonic front-end chips. Traditionally, RF photonic chips (a) are designed for one dedicated application.  More recently, these chips become programmable, enabling the reconfigurability for different functionalities (b). In this work (c), multiple microwave photonic functions are cascaded to achieve the first simultaneous multi-functional integrated RF photonic front-end chip.}
\label{fig1}
\end{figure}

To date, most reported integrated MWP front-ends are designed for specific applications, as illustrated in Fig.~\ref{fig1}(a), with a focus on singular functionalities such as filtering \cite{fandino2017monolithic,liu2020integrated,tao2021hybrid,wei2023ultrahigh,10255111}, phase shifting \cite{mckay2019brillouin,chew2022inline}, frequency measurement \cite{tao2022fully,ding2023simultaneous,10197393}, and beamforming \cite{zhuang2007single,pan2020microwave,shi2023compact,martinez2024ultrabroadband,r8-liu2024continuously}. However, as the demand for multifunctional integration and dynamic adaptability in RF systems grows, conventional single-function designs are increasingly inadequate for practical requirements. In response, programmable MWP front-ends capable of performing multiple functions have emerged in recent years \cite{zhou2020self,bogaerts2020programmable,r5-daulay2022ultrahigh,r10-liu2023linearized,xie2024towards,perez2024general}, as shown in Fig.~\ref{fig1}(b). These front-ends can be reconfigured to perform various tasks, enhancing the versatility and competitiveness of integrated MWP front-ends. 

Nevertheless, current programmable RF photonic chips are typically limited to executing only one function at a time. This limitation contrasts sharply with traditional RF front-ends, which feature a cascaded chain of signal processing functions, such as filtering, phase shifting, and frequency conversion. To meet the demands of advanced RF applications in complex electromagnetic environments, integrated MWP front-ends must be capable of processing RF signals with multiple functions simultaneously.

In this work, we present, for the first time, an integrated MWP front-end capable of performing simultaneous cascaded functions with enhanced performances, as shown in Fig.~\ref{fig1}(c). In contrast to previous efforts \cite{xue2012all,chew2016distributed}, which relied on multiple modulators and discrete fiber-based components, our approach leverages a single intensity modulator integrated with a low-loss silicon nitride photonic chip to achieve cascaded functionalities with high performances. Furthermore, we implemented an on-chip linearization technique to improve the spurious-free dynamic range (SFDR) of the system. This work marks an important step toward the development of high-performance, simultaneous multi-functional integrated MWP front-ends for advanced RF applications.

\section*{Principle of Operation}

The proposed cascaded multi-functional integrated MWP front-end is illustrated in Fig.2. This front-end can simultaneously perform true-time delay and notch filtering at different frequency bands. Additionally, by controlling the phases and amplitudes of the optical carrier, the spurious-free dynamic range (SFDR) of the system can be improved. In an alternative mode of operation, this front-end also supports simultaneous phase shifting and notch filtering.

\begin{figure*}[ht!]
\centering
\includegraphics[width=0.85\linewidth]{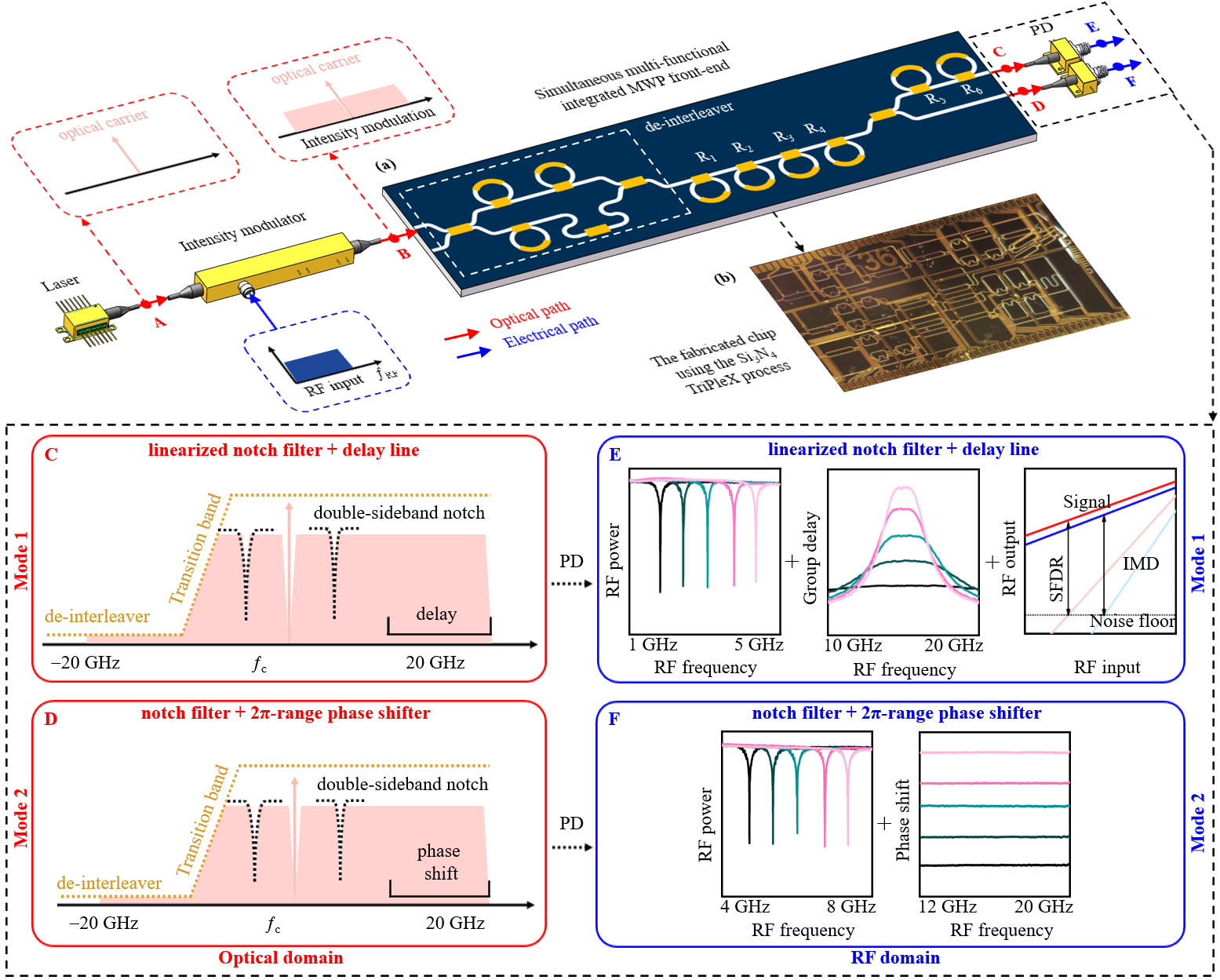}
\caption{The proposed cascaded multi-functional integrated MWP front-end. (a) Structure of the integrated MWP front-end chip, consisting of a spectral de-interleaver and cascaded all-pass ring resonators. (b) Photograph of the fabricated chip. (c) and (d) The de-interleaver filter out part of the lower sideband of the intensity modulated signal. (e) and (f) Four ring resonators in different coupling states are utilized for true time delay, phase shifting, and linearization.}
\label{fig2}
\end{figure*}



This MWP front-end consists of an on-chip tunable spectral de-interleaver \cite{luo2010high, r5-daulay2022ultrahigh, r10-liu2023linearized} and four cascaded all-pass ring resonators. This de-interleaver acts as a bandpass filter, selectively filtering out part of the lower sideband from an intensity-modulated signal. The four cascaded rings are tuned to different coupling states, providing distinct amplitude and phase responses to the remaining parts of the signal.


As shown in Fig.\ref{fig2}~(a), the de-interleaver is implemented using an asymmetric Mach–Zehnder interferometer (AMZI) that incorporates three micro-ring resonators (MRRs) \cite{guo2021versatile}, providing a flat-top complementary filter response with a free spectral range (FSR) of 160 GHz \cite{liu2021integrated}. This de-interleaver is followed by four cascaded MRRs with a FSR of 50~GHz (R\textsubscript{1} to R\textsubscript{4}) and two MRRs with a FSR of 80~GHz (R\textsubscript{5} and R\textsubscript{6}). The coupling ratios and round-trip phases of these MRRs are precisely controlled through thermo-optic tuning, offering tunability for various resonance states. By adjusting these voltages, the front-end chip can be configured and programmed into two operating modes.

The chip is fabricated using the Si\textsubscript{3}N\textsubscript{4} TriPleX process \cite{worhoff2015triplex,roeloffzen2018low} and is packaged by LioniX International BV. It demonstrates a low propagation loss of 0.1 dB/cm and a chip-to-fiber coupling loss of approximately 1.1 dB/facet.



The principle of operation to demonstrate the cascaded multi-functionality is as follows: the first mode of operation concerns the cascaded functions of notch filter and true time delay. This is illustrated in Fig.~\ref{fig2}(c) and Fig.~\ref{fig2}(e).  Here We symmetrically placed two resonances from two rings (R\textsubscript{1} and R\textsubscript{2}) on both sides of the optical carrier with equal notch rejection, one ring is set at an under-coupled (UC) state and the other is at an over-coupled (OC) state. This creates a \(\pi\) phase difference between both sidebands at the appointed notch frequency, resulting in high-rejection double-sideband notch suppression when converted back to the RF domain. 

Resonances from two additional rings, R\textsubscript{3} and R\textsubscript{4}, are used for continuously tunable true time delay by setting them to the over-coupling state. This creates Lorentzian-shape group delay responses. Further tuning of the rings coupling and central frequencies will lead to flattened group delay band. It is important to note that the key for the cascading of the filter and delay function is the judicious choice of spectral bands to have dual-sideband modulation (lower frequency range) and single-sideband (upper frequency range). 

In addition to the cascaded function, we employ two more ring resonators,  R\textsubscript{5} and R\textsubscript{6}, to tailor the amplitude and phase of the optical carrier for the purpose of linearization, effectively suppressing third-order intermodulation distortion (IMD3) and thereby increasing the spurious-free dynamic range (SFDR) of the system. The linearization condition is \cite{wei2024programmable}:

\begin{equation}
T\cdot \cos\theta=-\frac{1}{3}
\end{equation}
where \textit{T} is the power transmission of the optical carrier and $\theta$ is the phase shift to the optical carrier. 


For the second mode of operation, we cascaded the notch filtering with tunable phase shifting, as shown in Fig.~\ref{fig2}(d) and Fig.~\ref{fig2}(f). Here, the rings R\textsubscript{1} and R\textsubscript{2} are kept for achieving notch filtering, while rings R\textsubscript{3} and R\textsubscript{4} are configured to the OC state to introduce tunable phase shift to the optical carrier, thereby achieving the function of MWP phase shifter for the higher frequency range of the signal of interest.

\section*{Results}


\subsection*{Mode 1: Simultaneous linearized notch filtering and true time delay}

\begin{figure*}[t!]
\centering
\includegraphics[width=0.85\linewidth]{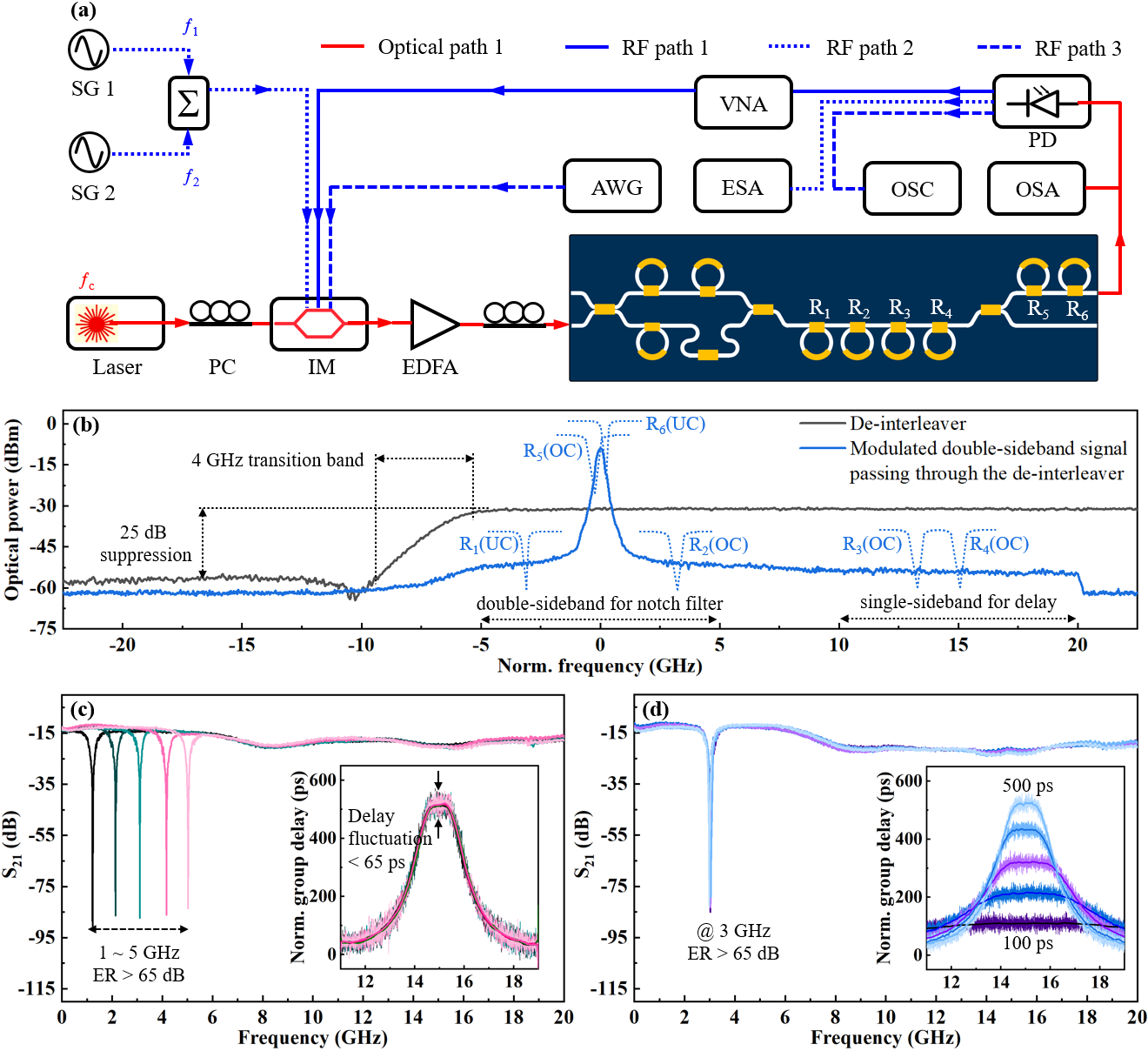}
\caption{Experimental setup and measurement results of simultaneous linearized notch filtering and true time delay. (a) Experimental setup. (b) Optical spectral responses of de-interleaver and the modulated double-sideband signal passing through the de-interleaver on the OSA. Amplitude and delay measurements of (c) varying notch positions with fixed group delay and (d) fixed notch position with varying group delays.}
\label{fig3}
\end{figure*}

The experimental setup for simultaneous linearized notch filtering and true time delay is depicted in Fig.~\ref{fig3}(a). The optical carrier at 1550~nm from a continuous-wave laser source (Pure Photonics PPCL550) is injected into a Mach-Zehnder intensity modulator (IM, Thorlabs LNA6213). It is modulated by an RF signal from the vector network analyzer (VNA, Keysight P9375A), signal generators (SG, Wiltron 69147A and HP 8672A), or arbitrary waveform generator (AWG, Keysight M8195A), depending on the various experiments. We use an erbium-doped fiber amplifier (EDFA, Amonics AEDFA-13)  to compensate for the insertion losses induced by the IM. Subsequently, the generated double-sideband intensity-modulated signal is processed by the MWP front-end chip. Following this, the signal is directed to a photo detector (PD, EMCORE 2522B) for conversion into the RF domain, while a portion of the optical signal is sent to an optical spectrum analyzer (OSA, Finisar 1500s) for optical spectrum analysis. The amplitude, phase, delay, S21 and other parameters of the RF signal processed by the MWP front-end chip are measured using electrical spectrum analyzer (ESA, Keysight N9000B), high-speed oscilloscope (OSC, Keysight UXR0134A) or VNA, for the various experiments discussed below.

The measured frequency responses of the de-interleaver along with the output intensity-modulated double-sideband signal shown illustrated in Fig.~\ref{fig3}(b). The de-interleaver provides a suppression of 25 dB with a 4-GHz (2.5\% of FSR) transition band and the modulated double-sideband signal undergoes filtering to remove the lower sideband signal ranging from -20 GHz to -5 GHz offset frequency from the optical carrier. The function of simultaneous double-sideband notch filtering and true time delay is subsequently achieved by tuning R\textsubscript{1}- R\textsubscript{4}. 

The magnitude response of the notch filter was measured using the VNA (indicated as RF path 1 in Fig.~\ref{fig3}(a)). The measurements encompassed varying notch positions with fixed group delays (Fig.~\ref{fig3}(c)), as well as fixed notch position with varying group delays (Fig.~\ref{fig3}(d)). As depicted in Fig.~\ref{fig3}(c), frequency tuning of the notch filter from 1 to 5 GHz was achieved while maintaining a notch rejection \textgreater 65 dB. The average 3-dB bandwidth of the notch filter is approximately 300 MHz. At the same time, we achieved true time delay response of  500 ± 3 ps with an average bandwidth of about 1.4 GHz, centered around 15 GHz. The time delay fluctuations of \textless 65 ps was achieved during the notch filter frequency tuning. 

Conversely, we measured the notch filter response when the group delay was tuned. Maximum group delay values of 500 ps, 400 ps, 300 ps, 200 ps, and 100 ps were achieved for respective delay bandwidth settings of approximately 1.4 GHz, 1.9 GHz, 2.3 GHz, 3.4 GHz and 6.2 GHz as depicted in Fig.~\ref{fig3}(d). Simultaneously, high-extinction (\textgreater 65 dB) notch filter centered at the frequency of 3~GHz was achieved.

\begin{figure}[t!]
\centering
\includegraphics[width=0.97\linewidth]{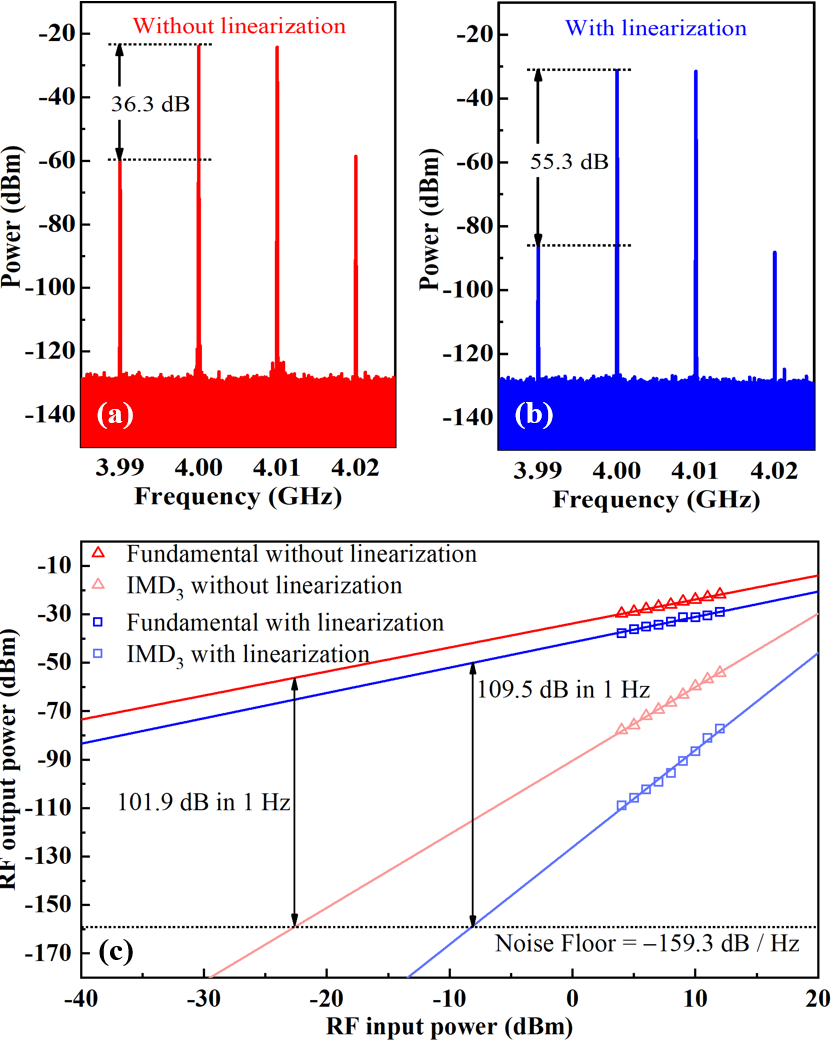}
\caption{Two-tone measurement results. Measured electrical spectrum of the link (a) without linearization and (b) with linearization. (c) Measured SFDR of the link without and with linearization.}
\label{fig4}
\end{figure}

\begin{figure}[t!]
\centering
\includegraphics[width=0.97\linewidth]{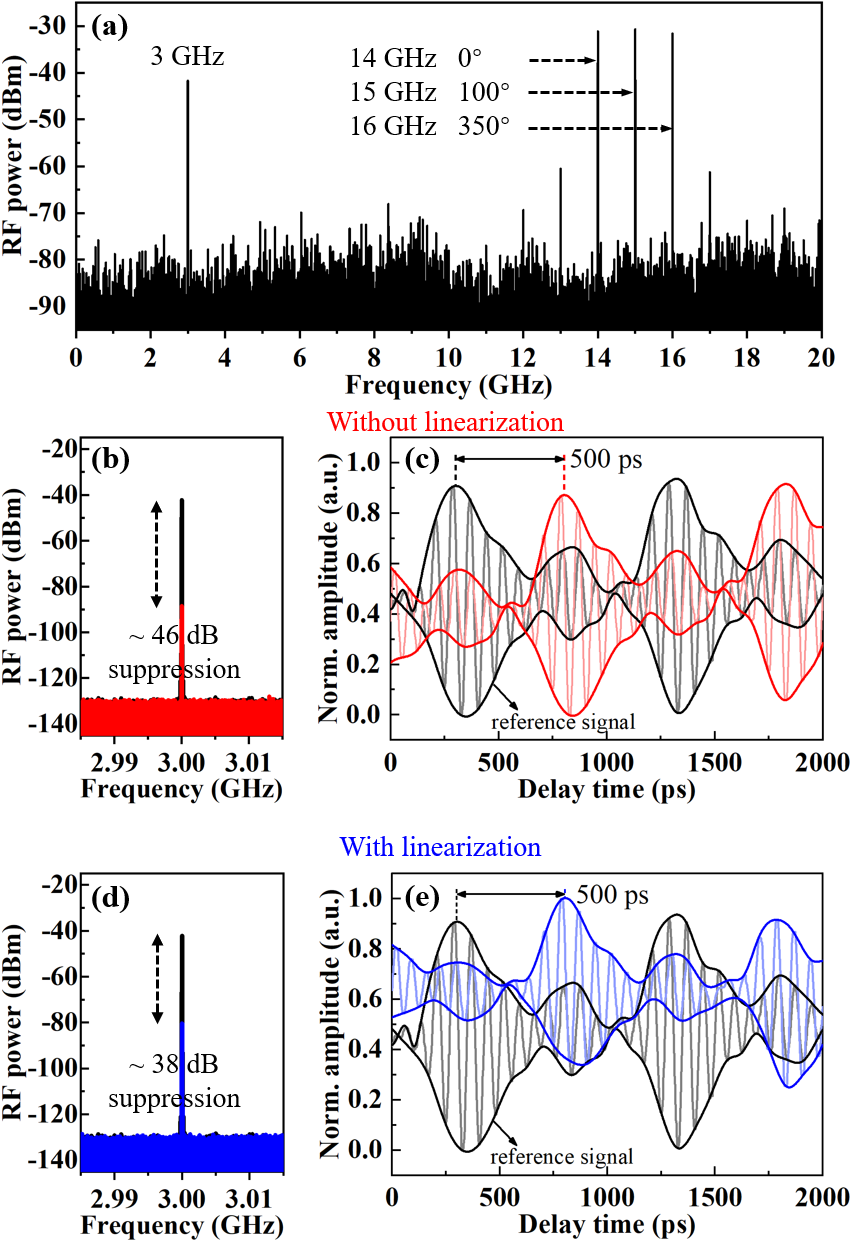}
\caption{System experiments. (a) Signals used for assessing both filtering and delay.  The 3 GHz signal generated by a signal generator emulates unwanted interference that should be filtered out. The broadband signal of interest is emulated by a combination of tones from 14 GHz to 16 GHz with different initial phases. (b) The noise suppression of the filter without linearization. (c) The time-domain delay measurement result without linearization. (d) The noise suppression of the the filter with linearization. (e) The time-domain delay measurement result with linearization.}
\label{fig5}
\end{figure}

It is critical to show that the proposed simultaneous cascaded signal processing can also achieve high RF system performance. For this reason, we carry out linearization experiments on the cascaded filter-time delay system. This is done through two-tone analysis, with the notch frequency fixed at 3 GHz and the group delay fixed at 500 ps (RF path 2 in Fig.~\ref{fig3}(a)). Two RF signals at frequencies of 4.00 GHz and 4.01 GHz and power of 8 dBm are generated using separate signal generators, combined via an RF combiner to form a two-tone signal. Subsequently, this two-tone signal is modulated onto the optical carrier using an IM. Due to the nonlinear characteristic of the IM, IMD3 components are generated at frequencies of 3.99 GHz and 4.02 GHz when the input two-tone signals are at frequencies of 4.00 GHz and 4.01 GHz as depicted in Fig.~\ref{fig4}(a). 

By properly phase shifting and suppressing the optical carrier using two MRRs (R\textsubscript{5} and R\textsubscript{6}), it is possible to suppress the three primary contributors of IMD3. We subsequently measured enhancement in the fundamental to IMD3 ratio from an initial value of  36.3 dB (Fig.~\ref{fig4}(a)) to 55.3 dB (Fig.~\ref{fig4}(b)). This translates into enhancement of the spurious-free dynamic range (SFDR), as illustrated in Fig.~\ref{fig4}(c). Here, we measured a shot-noise limited noise floor of -159.3 dBm/Hz, and SFDR enhancement from 101.9 dB to 109.5 dB within a bandwidth of 1 Hz.

\begin{figure*}[ht!]
\centering
\includegraphics[width=0.85\linewidth]{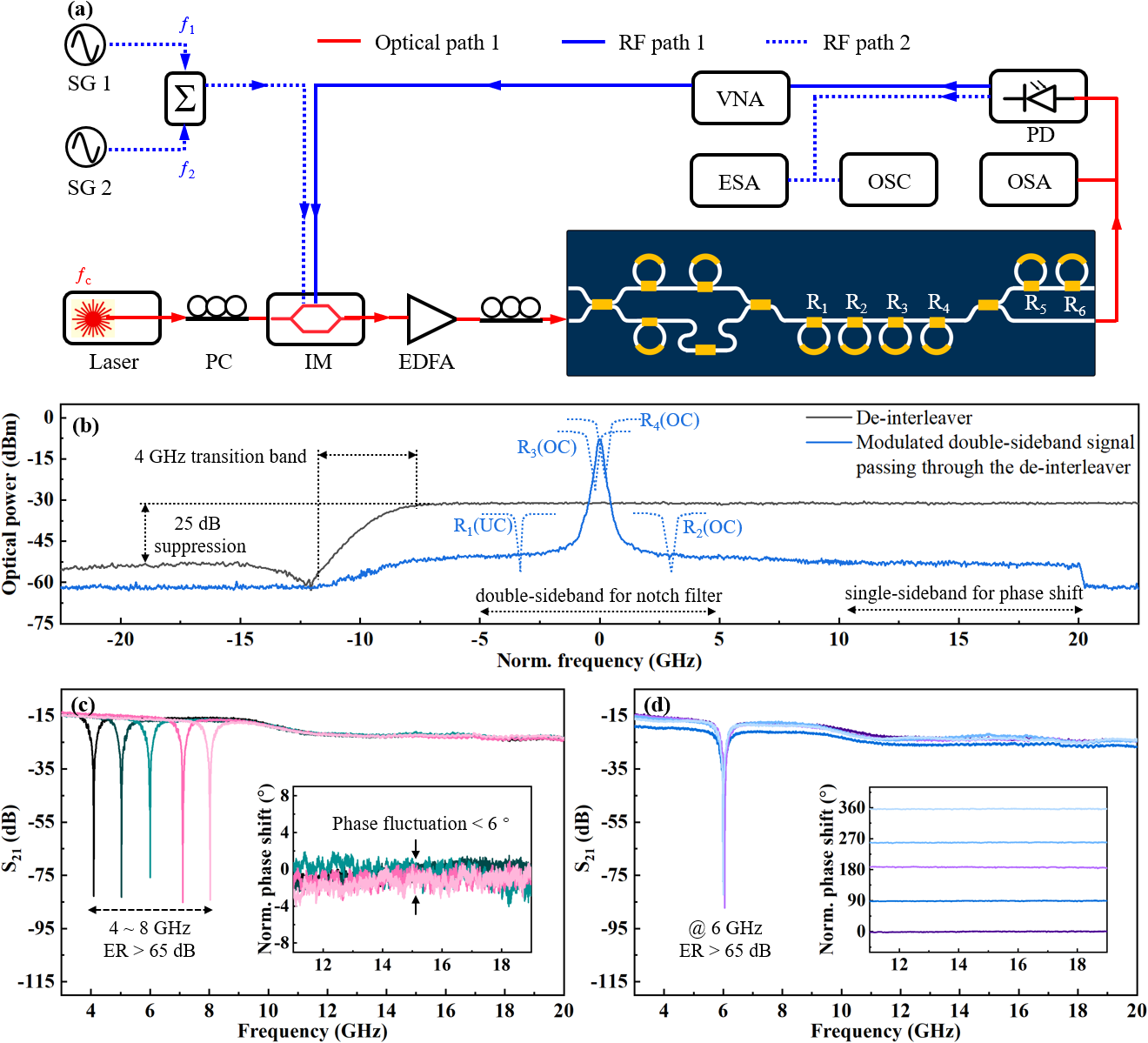}
\caption{Experimental setup and measurement results of simultaneous notch filtering and phase shifting. (a) Experimental setup. (b) Optical spectral responses of de-interleaver and the modulated double-sideband signal passing through the de-interleaver on the OSA. Amplitude and phase shift measurements of (c) varying notch positions with fixed phase shift and (d) fixed notch position with varying phase shifts.}
\label{fig6}
\end{figure*}

We further validate the cascaded functionalities through a system experiment. We emulate an interfering signal that requires filtering at the frequency of 3 GHz. Simultaneously, we emulate a signal of interest using an arbitrary waveform generator (AWG, Keysight M9502A). The signal spans the frequency of 14 GHz to 16 GHz with slower periodic envelope of 1 ns. The  measurement results are shown in Fig.~\ref{fig5}(b) to Fig.~\ref{fig5}(e). The interfering signal at the frequency of 3 GHz was suppressed by 46 dB (Fig.~\ref{fig5}(b)) while the signal of interest was delayed by 500 ps without any appreciable signal distortion (Fig.~\ref{fig5}(c)). 

Fig.~\ref{fig5}(d) and Fig.~\ref{fig5}(e)  depict the system performance When linearization is employed.  Although there is a slight decrease in the effectiveness of unwanted signal suppression compared to the link without linearization, high suppression of 38 dB can still be achieved. The signal delay, on the other hand, suffered from signal distortion, as depicted in Fig.~\ref{fig5}(e). We believe that this is primarily attributed to thermal cross-talk between the heaters as well as power and phase variations of the optical carrier.

\subsection*{Mode 2: Simultaneous notch filtering and phase shifting}

The MWP front-end chip can also be reconfigured to simultaneously implement notch filtering and phase shifting. The experimental setup is depicted in Fig.~\ref{fig6}(a), following a similar procedure as that for the implementation of Mode 1. The de-interleaver is employed to selectively filter out a portion of the lower sideband, R\textsubscript{1} and R\textsubscript{2} are utilized for double-sideband notch filtering, as depicted in Fig.~\ref{fig6}(b). Simultaneously,  R\textsubscript{3} and R\textsubscript{3} are tuned to the over-coupling state to introduce more than 2\(\pi\)-range tunable phase shift to the optical carrier to achieve the MWP phase shifter function. To reduce the power variation and power loss at different phase shift, the notches of the two OC MRRs are tuned to be close to each other and carrier suppression of \textless 1 dB can be achieved.

The amplitude and phase responses of the system were measured using the VNA (RF path 1) illustrated in Fig.~\ref{fig6}(a). The measurements included varying notch positions with a fixed phase shift (Fig.~\ref{fig6}(c)), as well as fixed notch position with varying phase shifts (Fig.~\ref{fig6}(d)). As can be seen in Fig.~\ref{fig6}(c), the notch position is tuned from 4 GHz to 8 GHz continuously and the suppression is always \textgreater 65 dB, whereas the phase shifter exhibits a fluctuation \textless 6°. Alternatively, as depicted in Fig.~\ref{fig6}(d), the notch position is fixed at 6 GHz and 0° \textasciitilde 360° phase shifting is achieved in frequency range of 12 GHz \textasciitilde 20 GHz. The maximum measurement frequency is constrained by the PD (EMCORE 2522B 20 GHz) and VNA (Keysight P9375A 26.5 GHz). In principle, the operational bandwidth of the phase shifter can reach approximately 50 GHz, as limited by the FSR of the ring resonators.

\begin{figure}[t!]
\centering
\includegraphics[width=0.95\linewidth]{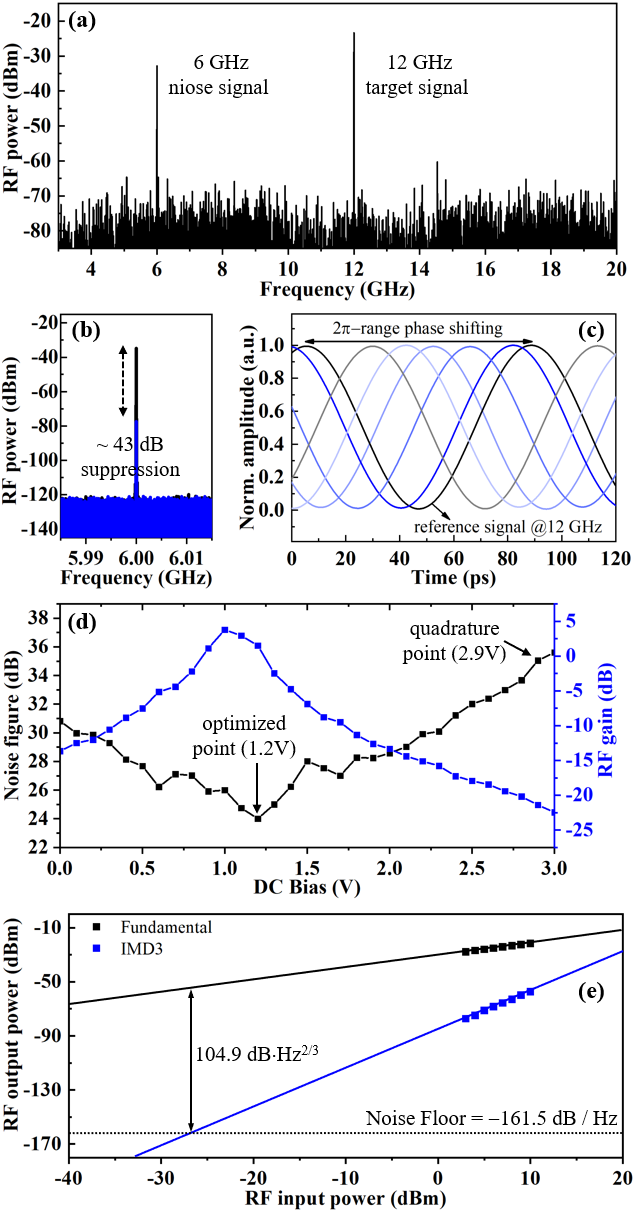}
\caption{System experiments for Mode 2: filtering and phase-shifting.(a) The spectrum of the input RF signal. (b) The interference suppression of the filter (c) 2\(\pi\)-range Phase shift in time domain of the 12 GHz signal of interest. (d) The link gain and noise figure of the system when tuning the bias voltage of the IM. (e) SFDR of the system after IM bias voltage optimization.}
\label{fig7}
\end{figure}

Finally, we carry out system experiments for Mode 2. We emulated unwanted interference  at 6 GHz and a signal of interest at 12 GHz, as shown in Fig.~\ref{fig7}(a). When the filter was activated,  interference suppression of  more than 43 dB can be achieved (Fig.~\ref{fig7}(b). At the same time, tunable phase shifting was implemented on the signal of interest at 12 GHz, covering the full 2\(\pi\)-range (Fig.~\ref{fig7}(c)).

For this mode of operation, we employed the low-biasing technique \cite{ackerman2008rf,urick2009analysis} to improve the RF gain and noise figure of the system. This technique involves adjusting the biasing point of the IM in conjunction of increasing the input power to the modulator to simultaneously achieve high link gain and low noise figure. By carefully tuning the bias voltage, significant improvements in system performance were observed. As shown in Fig.~\ref{fig7}(d), the RF gain can be increased from -22 dB to 3 dB when the IM is biased from the quadrature point (2.9 V) to an optimized point (1.2 V). This increase in RF gain is attributed to the enhanced modulation efficiency at the optimized bias point. Simultaneously, the noise figure of the system decreases from 35 dB to 24 dB. This reduction in noise figure is crucial for improving the overall sensitivity and signal-to-noise ratio of the system. The measured SFDR of the system was 104.9 dB·Hz\textsuperscript{2/3}, as shown in Fig.~\ref{fig7}(e).

\section*{Discussion and Conclusion}

In this paper, we introduced a novel programmable integrated MWP front-end capable of performing multiple simultaneous functions with enhanced RF performance metrics. This integrated front-end can operate in two distinct modes combining notch filtering and true time delay or phase shifting. The RF performance of the system is also enhanced through the utilization of a novel linearization technology and a well-established low-biasing techniques. The implementation of such multifunctionality within a single integrated platform marks a significant advancement in the field of MWP front-ends, offering promising advanced applications in real RF environments

With this work, we break that technological barrier that has long prevented the demonstration of simultaneous cascaded functions in integrated microwave photonics. The flexibility of tailoring modulation sidebands given by the spectral de-interleaver is key for this breakthrough. While this work uses a ring-loaded MZI as the de-interleaver, we believe that optimization of the de-interleaver structure will be critical to cascade more functions along with performance enhancement.  

Our current demonstration currently still suffers from thermal crosstalks and non-ideal tuning mechanism. Employing piezo or stress-based tuning can yield significant improvements. 

From the integration perspectives, our technique is compatible with current developments of heterogeneous integration, for example with on-chip laser sources \cite{zhou2023prospects}, amplifiers \cite{liu2022photonic}, modulators \cite{r3-feng2024integrated,wei2024programmable}, photodetectors \cite{chen2024design} in various emerging platforms such as thin-film lithium niobate.

\section*{acknowledgments}
The authors acknowledge funding from the European Research Council Consolidator Grant (101043229 TRIFFIC), Nederlandse Organisatie voor Wetenschappelijk Onderzoek (NWO) Start Up (740.018.021), China Scholarship Council program (202306090084).




\bibliography{ref}
\bibliographystyle{ieeetr}

\end{document}